\documentclass[sn-basic, Numbered]{sn-jnl}

\usepackage{graphicx}%
\usepackage{multirow}%
\usepackage{amsmath,amssymb,amsfonts}%
\usepackage{amsthm}%
\usepackage{mathrsfs}%
\usepackage[title]{appendix}%
\usepackage{xcolor}%
\usepackage{todonotes}
\usepackage{outlines}
\usepackage{verbatim}
\usepackage{textcomp}%
\usepackage{manyfoot}%
\usepackage{booktabs}%
\usepackage{algorithm}%
\usepackage{algorithmicx}%
\usepackage{algpseudocode}%
\usepackage{listings}%
\usepackage{bm}
\usepackage{bbm}

\begin{document}
\title{Assessing the impact of variance heterogeneity and misspecification in mixed-effects location-scale models}


\author*[1,2]{\fnm{Vincent} \sur{Jeanselme}}
\author*[1,3]{\fnm{Marco} \sur{Palma}}
\author[1]{\fnm{Jessica K} \sur{Barrett}}\email{}
\affil[1]{\orgdiv{MRC Biostatistics Unit}, \orgname{University of Cambridge}, \orgaddress{\country{UK}}}
\affil[2]{\orgdiv{Department of Biomedical Informatics}, \orgname{Columbia University}, \orgaddress{\country{USA}}}
\affil[3]{\orgdiv{Population, Policy and Practice Research and Teaching Department}, \orgname{UCL Great Ormond Street Institute of Child Health}, \orgaddress{\country{UK}}}


\abstract{
\textbf{Purpose:} Linear Mixed Model (LMM) is a common statistical approach to model the relation between exposure and outcome while capturing individual variability through random effects. However, this model assumes the homogeneity of the error term's variance. Breaking this assumption, known as homoscedasticity, can bias estimates and, consequently, may change a study's conclusions. If this assumption is unmet, the mixed-effect location-scale model (MELSM) offers a solution to account for within-individual variability.

\textbf{Methods:} Our work explores how LMMs and MELSMs behave when the homoscedasticity assumption is not met. Further, we study how misspecification affects inference for MELSM. To this aim, we propose a simulation study with longitudinal data and evaluate the estimates' bias and coverage.

\textbf{Results:} Our simulations show that neglecting heteroscedasticity in LMMs leads to loss of coverage for the estimated coefficients and biases the estimates of the standard deviations of the random effects. In MELSMs, scale misspecification does not bias the location model, but location misspecification alters the scale estimates.

\textbf{Conclusion:} Our simulation study illustrates the importance of modelling heteroscedasticity, with potential implications beyond mixed effect models, for generalised linear mixed models for non-normal outcomes and joint models with survival data.
}

\maketitle

\section{Background}

Longitudinal studies are critical to answering research questions in many fields because of their ability to highlight trends and changes over time. In medicine, these analyses are fundamental to capturing the evolution of a disease through different stages of life, as well as measuring the changes in treatments' effects. In social sciences, they are used to track the evolution of socioeconomic phenomena, such as the impact of where one lives on economic outcomes~\citep{wilson2012truly}, and the associated impact of policies on these outcomes.

In the analysis of longitudinal data, linear mixed models (LMMs) emerged as a popular option to model repeated measures for the same statistical unit, such as individuals or countries, because of their interpretability and the availability of implementations in multiple statistical software~\citep{verbeke1997linear, diggle2024longitudinal}.
The key advantage of LMMs is that one can model the population trajectory (i.e.\ the overall longitudinal trend) as a function of time and covariates. These coefficients, which apply to all statistical units, are known as fixed effects, while individual-specific coefficients, known as random effects, can capture the correlation between the observed measurements for the same unit and quantify the deviation with respect to the population trajectory. Random effects are assumed to be drawn from a common distribution (usually normal) with zero mean.

A common assumption made in the LMMs literature is \textit{homoscedasticity}: the residual variance is assumed constant across all observations and covariates. In other words, the effect of unobserved covariates, which is captured by the error term, contributes to variation in equal ways for all units. While some options to validate this assumption exist (e.g.~Levene's test~\citep{levene1960robust} for homogeneity of variance, or fitted-residual plots to graphically assess the absence of unexplained patterns), these are not commonly reported in practice. 
While heteroscedasticity does not necessarily bias fixed effect estimates, ignoring this heterogeneity can impact variance estimates and, consequently, result in Type I and II errors~\citep{rosopa2013managing, rosopa2019conditional}. 

Modelling variance heterogeneity at the individual level is becoming increasingly of interest in many clinical settings, as a way to quantify important population and individual characteristics in addition to average trends over time. 
For instance, in cardiovascular research, the notion of blood pressure variability~\citep{rothwell2010prognostic, parati2005blood} is becoming more commonly used as a risk factor for stroke and other events, as large fluctuations over time might be indicative of arterial health and other issues. Recent work has been published stating the importance of such concepts in cardiovascular studies \citep{stevens2019utility}. This could lead to the discovery of new treatments that not only bring blood pressure into a normal range but also stabilise its variability.  
In some instances \citep{muntner2015visit, yano2017visit}, summary statistics, such as the standard deviation or the coefficient of variation, computed for each individual have been used as measures of variability. \citet{barrett2019estimating} showed that these summary statistics are affected by measurement error and lack precision, especially in lower-sample-size settings, common to many longitudinal studies. This observation evidences that summary statistics are not effective in modelling variance heterogeneity, therefore, more sophisticated statistical models are needed to capture this heterogeneity.  


The statistics literature offers model-based solutions for addressing the quantification of the individual variance. \citet{hedeker2008application, hedeker2012modeling} introduced a mixed-effect location-scale model (MELSM) to model both the location parameter (the mean) and the scale parameter (the residual standard deviation) as a function of covariates and random effects.
MELSM effectively incorporates variance heterogeneity in the LMM framework, while reducing the reliance on the assumption of homoscedasticity (effectively, as MELSM includes LMM as a special case, the model can be used to understand whether the assumption holds). 
The model was initially applied to quantify the within-subject heterogeneity in ecological momentary assessments of mood variation \citep{rast2012modeling}, but has since been applied to other domains such as cystic fibrosis \citep{palma2024demographic} and, outside biomedical applications, student learning~\citep{leckie2024mixed}. In terms of computing options, MELSM can be fitted in MLwiN \citep{rasbash2000user} or on Mplus, by specifying the location-scale model as a multilevel structural equation model \citep{mcneish2021specifying}. A more general modelling framework is provided in \textit{gamlss} \citep{gamlss} and its Bayesian counterpart (\textit{bamlss}, \citealp{umlauf2018bamlss}): a large range of outcome distributions is available, and all parameters of the distribution can be modelled using fixed and random effects (although not many examples explicitly model within-individual variability using random effects). These options fall under the umbrella of distributional regression models \citep{klein2024distributional}. 

Despite the existence of such solutions, modelling heteroscedasticity is not often done in practice. \citet{walters2018power} discussed how software availability and a lack of best practices in specifying the scale model may explain this lack of adoption. With recent user-friendly software developments such as~\cite{burkner2017brms} and \cite{gamlss}, only the second barrier remains. Further, methodological challenges exist when using MELSMs. For example, although the problem of power analysis and testing the homogeneity of variance in MELSM was the object of some recent studies \citep{walters2018power, williams2021reliability}, the relationship between location and scale is not fully understood, and how the misspecification of one affects the other. 

Our work aims to fill this gap by exploring how common modelling practices and associated misspecifications impact the estimates of interest. Through a comprehensive set of simulations, we explore how ignoring heteroscedasticity and misspecifying the location and scale models impact random and fixed effects estimations. Our work highlights how MELSM offers a robust framework for handling heteroscedasticity.

The paper is structured as follows. Section 2 introduces our notations and the studied MELSM. Section 3 describes the simulation setting. In Section 4, we study six common practices in specifying the scale model and explore how misspecifications impact MELSM's estimates. Finally, in Section 5, we discuss the insights from the simulations, their limitations, and potential further developments.

\section{Mixed-effect location-scale model}
Consider the normally-distributed outcome $y_{i, j} \in \mathbb{R}$ for subject $i \in [1, N]$ observed at encounter $j \in [1, n_i]$ sampled from a non-informative encounter process\footnote{When this assumption does not hold, one needs to model the encounter process as proposed in~\cite{gasparini2020mixed}.}, where $[a, b]$ denotes the interval of integers from $a$ to $b$, inclusive. The value $N$ is the number of subjects, and the value $n_i$ is the number of observations for subject $i$. A linear mixed effect model assumes the following relation between outcome and covariates:

$$y_{i, j} = x_{i,  j}^{y T} \beta^y + z_{i,  j}^{yT} u^y_i + \epsilon_{i, j}.$$

We take the errors $\epsilon_{i,j}$ to be normally distributed. This model implies that $y_{i,j} \sim \mathcal{N}(x_{i,  j}^{y T} \beta^y + z_{i,  j}^{yT} u^y_i, \omega^2)$, where the standard deviation $\omega$ in the normal distribution $\mathcal{N}$ is constant for all observations (homoscedasticity). The MELSM relaxes this assumption by specifying a model for the scale, which can depend on random effects, as well as covariates, i.e.:
%
$$\omega_{i,j}  = \exp(x_{i, j}^{\omega T} \beta^\omega + z_{i,j}^{\omega T} u^\omega_{i})$$
where $\beta^\omega$ specifies the contribution of the different covariates to the variance. We model the logarithm of the standard deviation to ensure non-negativity. As both location and scale present fixed effects (denoted by $\beta$) and random effects (denoted by $u_i$), we distinguish the different parameters using superscripts $\omega$ and $y$. 
Note that, for the normal distribution, the location and scale correspond to the traditional mean and standard deviation. In the remainder, we use the terms location and scale to distinguish the submodels' parameters.

The random effect vectors $u^y_i$ and $u^\omega_i$ are to be interpreted as \textit{subject-specific} deviations from the population mean and standard deviation, respectively. For ease of notation, random intercepts $u^y_{0, i}$ and $u^\omega_{0, i}$ are implicitly included by setting the first column of the corresponding covariate matrices to a vector of 1s, while random slopes $u^y_{1, i}$ and $u^\omega_{1, i}$ are obtained by setting the column to the vector of observation times. 
The random effects for the location and scale models are assumed to be normally distributed with zero mean, and a correlation between random effects can be estimated. For example, in the simple setting with just a random intercept for both the location and scale models, we can specify the correlation $\rho$ to capture the relationship between the random intercept for the mean and the random intercept for the variance:
\begin{equation}
   \begin{pmatrix}u_{i}^y\\
u_{i}^w
\end{pmatrix} \sim  \mathcal{N}
\begin{bmatrix}
\begin{pmatrix}
0\\
0
\end{pmatrix}\!\!,
\begin{pmatrix}
\sigma_y^{2}& \rho\sigma_{y}\sigma_\omega \\
\rho\sigma_{y}\sigma_{\omega}  & \sigma_{\omega}^2
\end{pmatrix}
\end{bmatrix} 
\label{eq:randomeffects_MELSMbasic}
\end{equation}
We denote the covariance matrix in \eqref{eq:randomeffects_MELSMbasic} as $\Sigma$.

\section{Simulation study}
This section describes the data generation and evaluation used in the proposed simulations to evaluate the robustness of MELSM to different misspecifications.

\subsection{Data generation}
\label{sec:generation}
We generate simulation data based on the Primary Biliary Cholangitis (PBC) dataset, publicly available in the R package \textit{survival} and commonly used as a test for longitudinal modelling. The dataset results from a clinical trial at Mayo Clinic studying the impact of D-penicillamine on PBC, an autoimmune condition affecting the liver. From the dataset, we extract four covariates at baseline: age, albumin, triglycerides (trig), and platelets.

Using these covariates, we generate a synthetic population of $N$ individuals by drawing from a normal distribution with the covariance defined by the previously extracted and standardised covariates at baseline. Note that $N$ is a parameter of the simulation and not the observed PBC population size, as one of our aims is to study how an increasing number of patients impacts our conclusions. Formally, we draw $x_{i, 0}$ -- where $i$ is the index associated with the patient and $0$, the baseline encounter -- from a multivariate normal distribution $\text{MVN}(0, \Sigma)$ with the entries of the covariance matrix $\Sigma$ given in Table~\ref{tab:correlation}.

\begin{table}[h!]
    \centering
    \begin{tabular}{|r|cccc|}
    \hline
              & age    & albumin & trig   & platelet \\
    \hline
    age       & 1.02   & -0.23   & 0.02   & -0.14    \\
    albumin   & -0.23  & 0.90    & -0.10  & 0.18     \\
    trig      & 0.02   & -0.10   & 1.00   & 0.10     \\
    platelet  & -0.14  & 0.18    & 0.10   & 0.89     \\
    \hline
    \end{tabular}
    \caption{Observed covariance matrix for baseline covariates in the PBC dataset.}
    \label{tab:correlation}
\end{table}
As we model longitudinal data, we then draw a number of encounters $J_i$ from a discrete uniform distribution for each patient $i$: 
$$J_i \sim \mathcal{U}\{1, 2 M - 1\}$$
where $\mathcal{U}(a, b)$ denotes a uniform distribution with lower and upper bounds $a$ and $b$, respectively, and $M$ is a parameter of the simulation study representing the average number of encounters per subject across the population. This encounter process reflects a regular schedule uninformative of the outcome of interest. For each encounter, we compute the age at this encounter as follows:
$$\text{age}_{i, j} = \text{age}_{i, 0} + j$$
and we use this covariate as the only time-varying variable while all other covariates remain constant over time. Unless otherwise specified, we consider a total of $N = 200$ patients with an average number of observations $M = 15$. 
%
Further, we simulate random intercepts for all individuals defined as in Equation~\ref{eq:randomeffects_MELSMbasic}.
We set $\rho = 0$, $\sigma_y = 1$ and $\sigma_\omega = 0.5$ to induce non-negligible variability in the simulated outcomes.

MELSM models a longitudinal outcome of interest. To this end, we generate the outcome as follows:
\begin{align*}
    y_{i, j} &= \beta^y \cdot x_{i, j} + u^y_{0,i} + \epsilon_{i,j} \tag{Location}\\
    \epsilon_{i,j} &\sim \mathcal{N}(0, \omega^2_{i,j})\\
    \omega_{i,j} &= \exp(\beta^\omega \cdot x_{i, j}+ u^\omega_{0,i}) \tag{Scale}
\end{align*}

In our simulations, we set the fixed effect to depend on two covariates for location and scale (to allow for a variety of plausible misspecification types), as summarised in Table~\ref{tab:effects}.

\begin{table}[h!]
    \centering
    \begin{tabular}{|r|cccc|}
    \hline
              & age    & albumin & trig   & platelet \\
    \hline
    $\beta^y$       & 0.5   & 0.5   & 0   & 0    \\
    $\beta^\omega$   & 0.8  & 0    & 0.8  & 0     \\
    \hline
    \end{tabular}
    \caption{Fixed effects for location and scale.}
    \label{tab:effects}
\end{table}

\subsection{Estimands of interest}
Our simulations explore the recovery of the fixed and random effects for both scale and location models. To this end, we report the box plots and the associated coverage (for 95\% credible intervals) for the estimated coefficients of interest across 500 repetitions of the simulation described in Section~\ref{sec:generation}.
Formally, the coverage of the parameter $\theta$ (in percent) is defined as follows:
$$\sum_s \mathbbm{1}(L(\hat{\theta}, \alpha) <\theta< U(\hat{\theta}, \alpha))$$
in which $\mathbbm{1}(c)$ denotes the indicator function, which equals 1 when the condition $c$ is satisfied and 0 otherwise; and $L$ and $U$ are the lower and upper bounds of the estimated $\hat{\theta}$ at the confidence level $\alpha$ computed as the quantiles of the posterior samples, meaning that for accurately estimated parameters and standard errors the coverage probability should be $\alpha$. Additional metrics (estimated credible interval width, the standard deviation of the estimates, and average estimated standard errors across the different replications) are presented in Appendix B to further support the insights of the conclusions presented below.

\subsection{Software}
All experiments are implemented in R with code made available on GitHub\footnote{\url{https://github.com/Jeanselme/IndividualVariability}}. We fit the MELSM models using the \textit{brms} package~\citep{burkner2017brms}, with default priors, two chains, and a burn-in of 1000 iterations for a total of 2000 iterations.

\section{Results}
This section describes common practices when using mixed-effect models and their impact on parameter' estimates. For each model, we report the formula in pseudo-R code for reproducibility.

\subsection{Practice 1: Ignoring heteroscedasticity}
By relying on LMM, practitioners typically assume homoscedasticity, thereby ignoring potential subject-specific differences in variance. In this first setting, we compare an LMM, assuming a constant scale, with the properly specified MELSM model.
The two models share the same (correct) specification for the location, with only the scale differing:
\begin{itemize}
    \item \textbf{MELSM} correctly models the scale.
    \begin{verbatim}
        y ~ age + albumin + (1|id)
        log(omega) ~ age + trig + (1|id)
    \end{verbatim}
    
    \item \textbf{LMM} assumes homoscedasticity.
    \begin{verbatim}
        y ~  age + albumin + (1|id)
        log(omega) ~ 1
    \end{verbatim}
\end{itemize}

\begin{figure}[hbtp]
    \centering
    \includegraphics[width=0.49\textwidth]{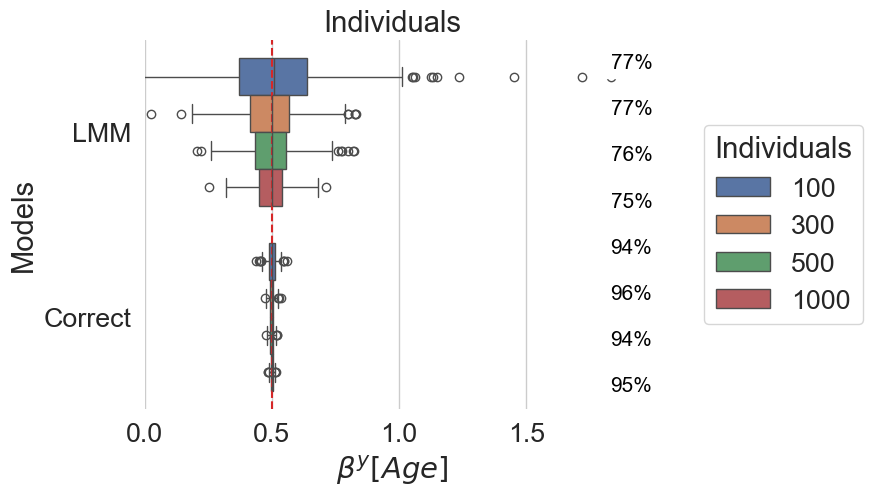}\hfill
    \includegraphics[width=0.49\textwidth]{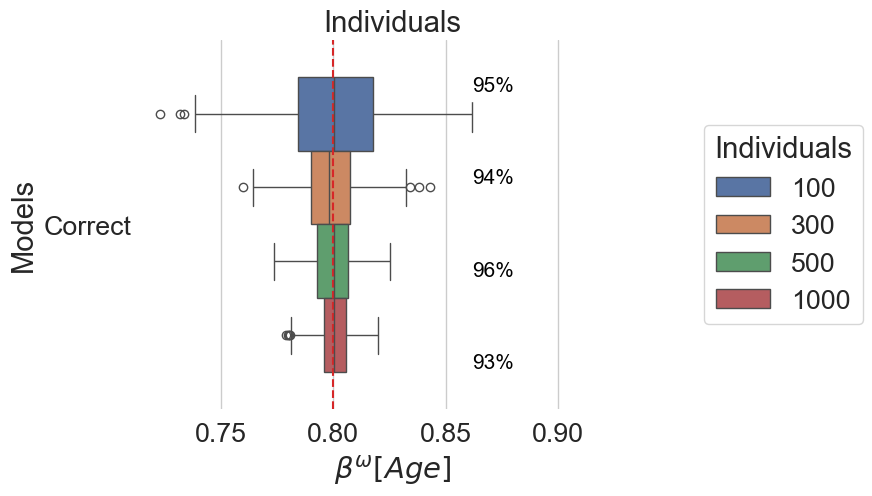}    
    \caption{Estimated age effect on location (left panel) and scale (right panel) for increasing number of individuals. For each model, coverage is reported as a percentage next to the corresponding boxplot.}
    \label{fig:s1:number}
\end{figure}

\begin{figure}[hbtp]
    \centering
    \includegraphics[width=0.49\textwidth]{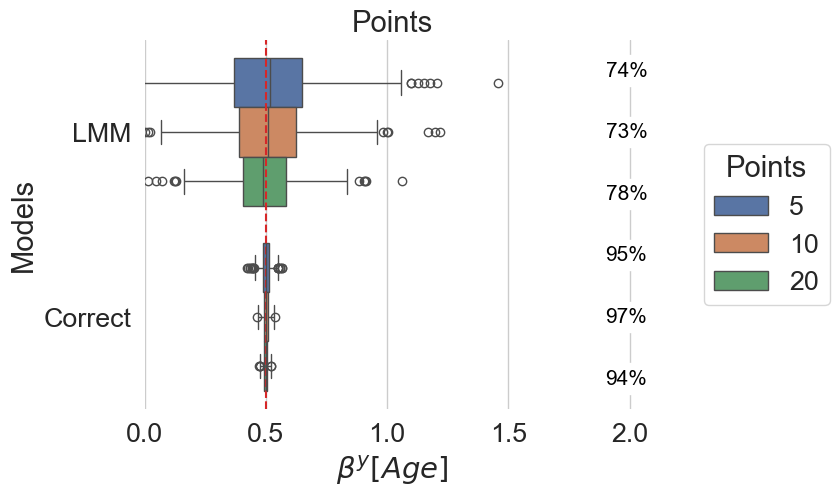}\hfill
    \includegraphics[width=0.49\textwidth]{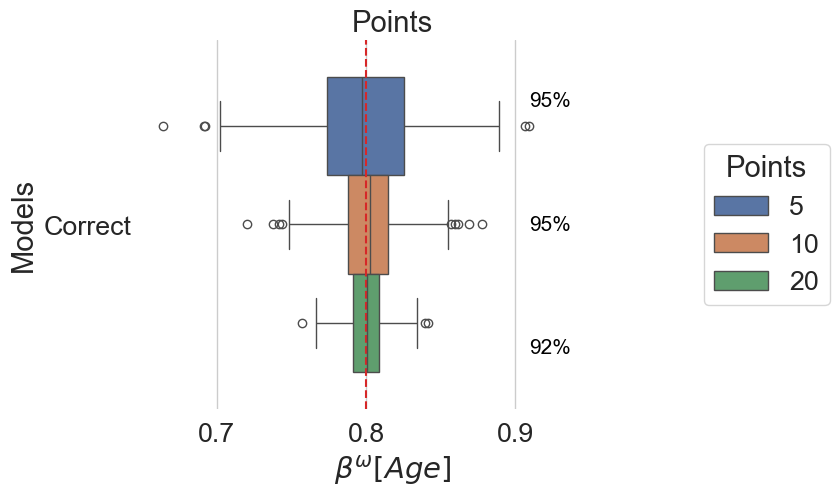}    
    \caption{Estimated age effect on location (left panel) and scale (right panel) for increasing average number of observations per individual. For each model, coverage is reported as a percentage next to the corresponding boxplot.}
    \label{fig:s1:points}
\end{figure}

In this first set of simulations, we investigate the effect of varying the number of patients and observations per patient, i.e., $N \in \{100, 300, 500, 1000\}$ and $M \in \{5, 10, 20\}$. These values were chosen to approximate different data amounts and frequencies as common in real longitudinal data settings. 
Figure~\ref{fig:s1:number} presents the estimated fixed effects associated with age in the location and scale models under varying numbers of individuals $N$ across 100 simulations. Similarly, Figure~\ref{fig:s1:points} presents the same results under varying average numbers of observations per patient.  In each figure, the red vertical line indicates the true value of the estimand. 

\paragraph{Scale misspecification impacts the precision of location estimates.} Under both analyses, Figures~\ref{fig:s1:number} and~\ref{fig:s1:points} show that the estimated age coefficients in the location model, although centered around the true value, display larger variance for LMM than MELSM across all sizes of the generated datasets. 

\paragraph{Coverage is low in misspecified settings.}
The coverage in the LMM is well below the nominal level of 95\%, indicating that the standard errors are underestimated. Consequently, the misspecification of the scale has a direct impact on the inference for the location parameters. 

\paragraph{Larger numbers of individuals and encounters reduce error.} Both models benefit from a larger number of individuals and encounters, as shown by reduced variance in the location estimates and reduced variance for MELSM scale estimates.

\subsection{Practice 2: Misspecifying location and scale in MELSM}
As the parametric form of either scale or location is rarely known in advance, we explore the influence of different types of misspecification (adding irrelevant covariates, neglecting relevant covariates or misspecifying the random effect structure) within MELSM. Specifically, we consider the following modelling of the generated data:
\begin{itemize}
    \item \textbf{Correct} is the correctly specified model.
    \begin{verbatim}
        y ~ age + albumin + (1|id),
        log(omega) ~ age + trig + (1|id)
    \end{verbatim}
    \item \textbf{All} considers all covariates for both models.
    \begin{verbatim}
        y ~ age + albumin + trig + platelet + (1|id)
        log(omega) ~ age + albumin + trig + platelet + (1|id)
    \end{verbatim}
    \item \textbf{Misspecified $y$} ignores age in the location model.
    \begin{verbatim}
        y ~ albumin + (1|id)
        log(omega) ~ age + trig + (1|id)
    \end{verbatim}
    \item \textbf{Misspecified $\omega$} does not consider age in the scale expression.
    \begin{verbatim}
        y ~ age + albumin + (1|id)
        log(omega) ~ trig + (1|id)
    \end{verbatim}
    \item \textbf{No $u^\omega$} ignores the random effect associated with the scale model.
    \begin{verbatim}
        y ~ age + albumin + (1|id)
        log(omega) ~ age + trig
    \end{verbatim}
    \item \textbf{No $u^\omega$ and misspecified $\omega$} combines the two previous scenarios, with age and random intercept ignored in the scale model.
    \begin{verbatim}
        y ~ age + albumin + (1|id)
        log(omega) ~ trig
    \end{verbatim}
\end{itemize}

\begin{figure}[hbtp]
    \centering
    \includegraphics[width=0.49\textwidth]{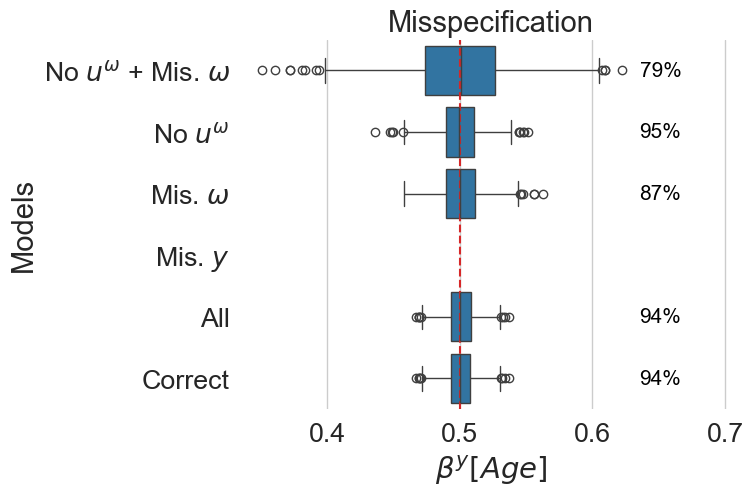}\hfill
    \includegraphics[width=0.49\textwidth]{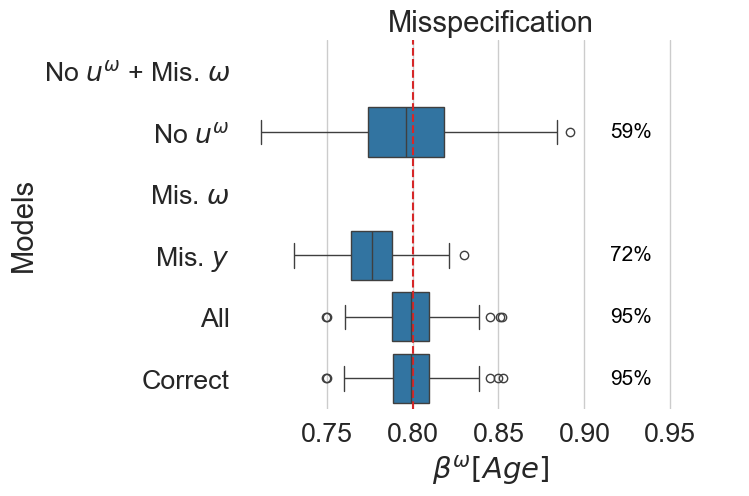}    
    \caption{Estimated age effect for location (left panel) and scale (right panel). For each model, coverage is reported as a percentage next to the corresponding boxplot.}
    \label{fig:s2:age}
\end{figure}

\begin{figure}[hbtp]
    \centering
    \includegraphics[width=0.49\textwidth]{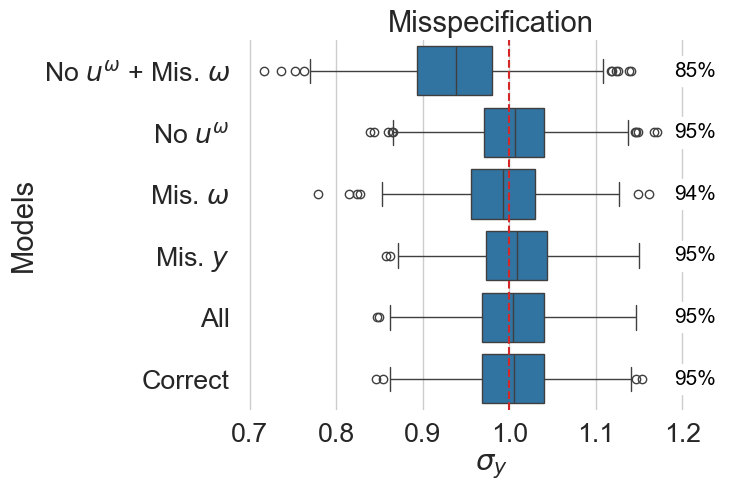}\hfill
    \includegraphics[width=0.49\textwidth]{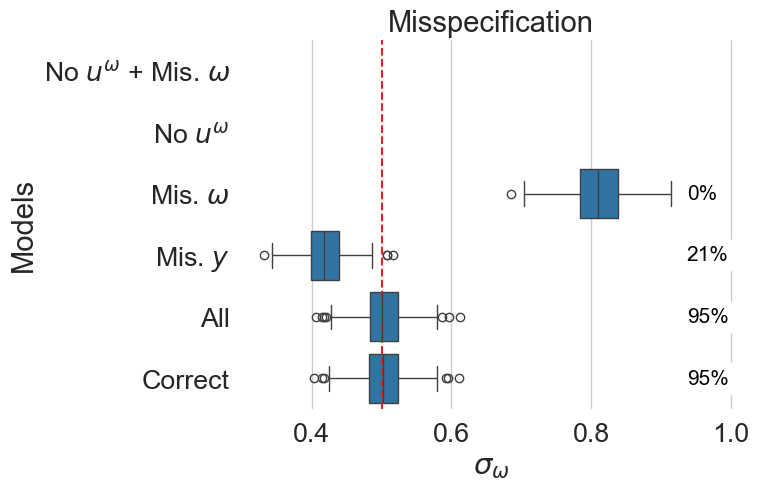}    
    \caption{Estimated standard deviation of random intercepts for location (left panel) and scale (right panel). For each model, coverage is reported as a percentage next to the corresponding boxplot.}
    \label{fig:s2:ri}
\end{figure}

Figure~\ref{fig:s2:age} presents the fixed effect estimates associated with the age covariate, while Figure~\ref{fig:s2:ri} describes the estimated standard deviation of the random intercept across the simulations. On the right, we display the coverage in percent over the different simulations. For consistency, we display all method specifications in these figures. Note that some do not estimate all quantities, hence the absence of some estimates.

\paragraph{Adding superfluous parameters does not invalidate inference.} In all figures, the correctly specified model (Correct) and the one considering all covariates (All) result in similar estimates. This indicates that, with sufficient data, all models that include the correct covariates as a special case will yield a correct inference. This is also confirmed by the other metrics in Appendix B. 
Note, however, that considering all covariates results in an increased number of parameters and associated computational time; the fitting procedure for the method considering all covariates is more than 50\% slower than the one for the correctly specified model.

\paragraph{Not all scale misspecification types affect the location in the same way.} The age effect in the location model is well estimated across all models, although the variance of the estimates increases as the scale model departs from the true one (as observed with LMM). The largest variance is observed for \textit{No $u^\omega$ + Mis. $\omega$}, where both a relevant covariate and the random intercept are missing in the scale model. The same model also shows, on average, lower estimated standard deviations for the location random intercept than the true value. 

\paragraph{Location misspecification impacts scale.} In the misspecified location model \textit{Mis. $y$}, the scale fixed effect estimate and random intercept are biased, despite presenting the right scale specification. As the location parameterisation does not capture the full variation of the data, the residual variance differs from the true value, and therefore, all scale estimates are biased (Appendix A provides a theoretical intuition of why this occurs).

However, the opposite only affects the variance of the estimates: when the scale is misspecified, as in \textit{Mis. $\omega$}, the location's age effect estimate remains unbiased. When the scale model ignores age (\textit{Mis. $\omega$}) or ignores the random intercept (\textit{No $u^\omega$}), the location effect presents a larger variance. Further, when both random intercept and misspecification of the fixed effect occur (\textit{No $u^\omega$ + Mis. $\omega$}), the variance more than doubles.

Therefore, location misspecification results in biased scale estimates, while the opposite only impacts the variance of the estimands.

\paragraph{There is a relationship between random effects and residual variance.} The estimated standard deviation of location random intercept can incur bias when the scale model is incorrectly specified (and similarly when a LMM is wrongly used in place of a MELSM). The standard deviation for the random effect of the scale model is biased in all considered scenarios unless the fixed effect structure for both location and scale is correctly specified.

\subsection{Practice 3: Ignoring non-linear time trend in location model}
Another common type of misspecification in everyday practice is the choice of a linear relationship between the outcome mean and the time variable when, in fact, a non-linear trend is present. We, therefore, explore how non-linearity would affect the estimates of interest. To this end, we alter the data generation mechanism 
by applying a sinusoidal transformation on age as follows:
$$y_i = \beta^y_{age}\sin(age_i) + \beta^y_{alb} albumin_i + u^y_i + \epsilon_i$$
while we keep a linear relationship with the scale parameter.
In this context, we compare the properly specified models with the common misspecification of ignoring the non-linearity (while the scale is correctly specified):
\begin{itemize}
    \item \textbf{Correct} consists of the properly specified model under this data generation process.
    \begin{verbatim}
          y ~ sin(age) + albumin + (1|id),
          log(omega) ~ age + trig + (1|id)
    \end{verbatim}
    \item \textbf{Non sinus} ignores the non-linearity on the covariates in the location.
    \begin{verbatim}
          y ~  age + albumin + (1|id)
          log(omega) ~ age + trig + (1|id)
    \end{verbatim}
\end{itemize}

\begin{figure}[hbtp]
    \centering
    \includegraphics[width=0.49\textwidth]{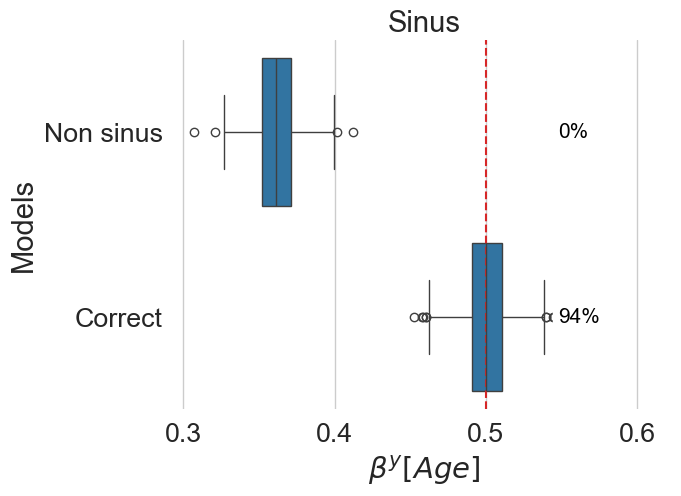}
    \includegraphics[width=0.49\textwidth]{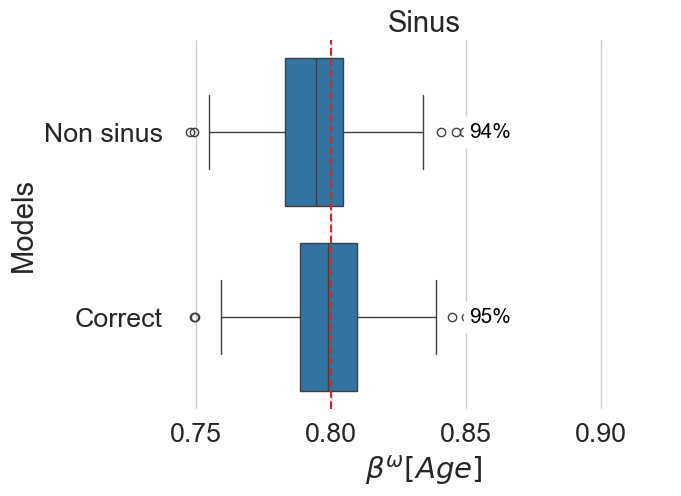}    
    \caption{Estimated age effect for location (left panel) and scale (right panel). For each model, coverage is reported as a percentage next to the corresponding boxplot.}
    \label{fig:s3:age}
\end{figure}

Figure~\ref{fig:s3:age} presents the fixed effect estimate associated with the age covariate. 

\paragraph{Non-linearity biases location fixed effects.} In this case, the misspecification of the functional form of age leads to a biased estimate of the effect of the covariate on the location model. As the scale relies on age and not a functional transformation, the nonlinearity in the location model has little impact on the scale submodel.

\subsection{Practice 4: Misspecifying the random effect structure}
One drawback of MELSM is that the computation becomes heavier as the number of random effects increases. For this reason, a practitioner might decide to have a more parsimonious model structure with only random intercepts for both location and scale models, even if random slopes might actually be justified in the analysis.

In the simulation settings discussed so far, we have generated the data assuming only a random intercept for the location and a random intercept for the scale, while in this setting, we alter the data generative process by adding an uncorrelated random slope for both location and scale. We then explore how different misspecifications may alter the estimates of the fixed effect. In these experiments, the generative process of the scale and location is as follows:

\begin{align*}
    y_{i, j} &= \beta^y \cdot x_{i, j} + u^y_{0, i} + \bm{u^y_{age, i} \text{age}_{i,j}} + \epsilon_{i,j}\\
    \epsilon_{i,j} &\sim \mathcal{N}(0, \omega^2_{i,j})\\
    \omega_{i,j} &= \exp(\beta^\omega \cdot x_{i, j} + u^\omega_{0, i} + \bm{u^\omega_{age, i} \text{age}_{i,j}})
\end{align*}

in which both $u^y_{age, i}$ and $u^\omega_{age, i}$ are drawn from normal distributions: $\mathcal{N}(0, 1)$ resp. $\mathcal{N}(0, 0.5^2)$. 
In this context, we consider three models that properly control on the correct covariates but specify different random effects:
\begin{itemize}
    \item \textbf{Correct} models both random slopes.
    \begin{verbatim}
          y ~ age + albumin + (1 + age|id),
          log(omega) ~ age + trig + (1 + age|id)
    \end{verbatim}
    \item \textbf{No $u_{age}^\omega$} ignores the scale random slope.
    \begin{verbatim}
          y ~ age + albumin + (1 + age|id),
          log(omega) ~ age + trig + (1|id)
    \end{verbatim}
    \item \textbf{No slopes} ignores both random slopes.
    \begin{verbatim}
          y ~  age + albumin + (1|id),
          log(omega) ~ age + trig + (1|id)
    \end{verbatim}
\end{itemize}

\begin{figure}[hbtp]
    \centering
    \includegraphics[width=0.49\textwidth]{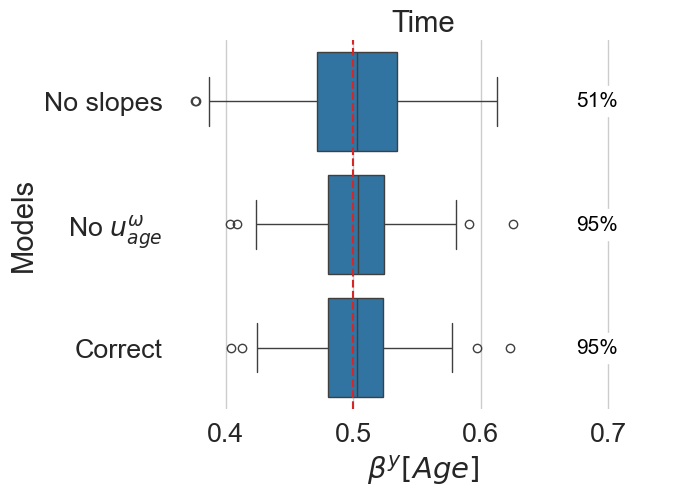}
    \includegraphics[width=0.49\textwidth]{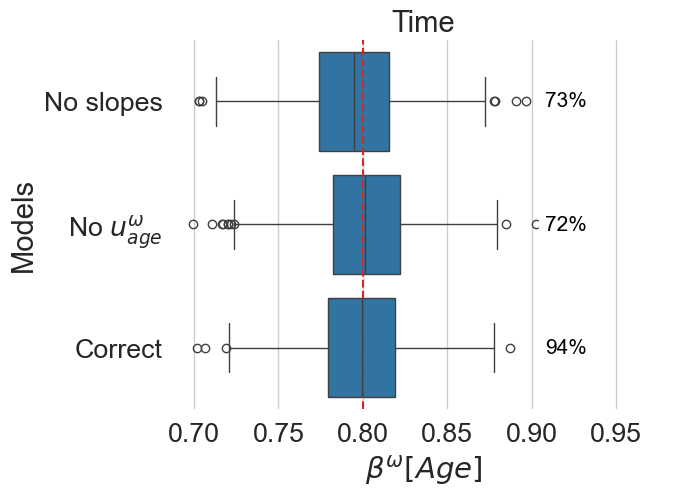}    
    \caption{Estimated age effect for location (left panel) and scale (right panel). For each model, coverage is reported as a percentage next to the corresponding boxplot.}
    \label{fig:s4:age}
\end{figure}

\begin{figure}[hbtp]
    \centering
    \includegraphics[width=0.49\textwidth]{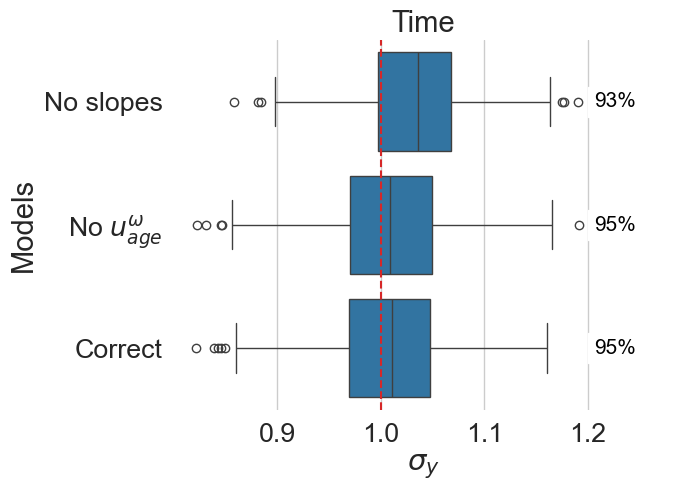}
    \includegraphics[width=0.49\textwidth]{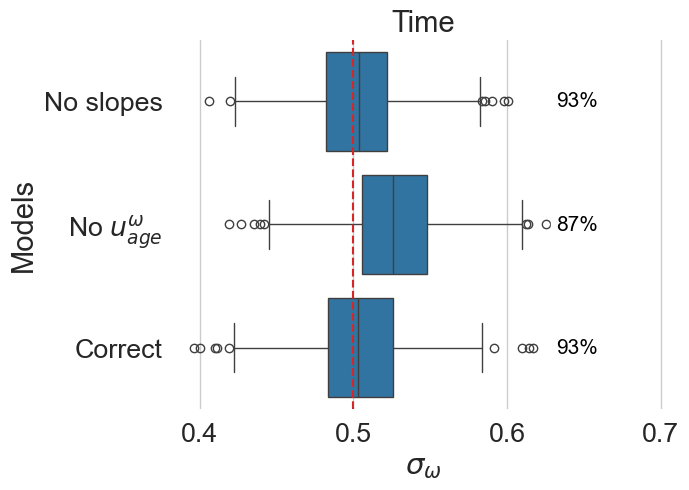}    
    \caption{Estimated standard deviation of random intercepts for location (left panel) and scale (right panel). For each model, coverage is reported as a percentage next to the corresponding boxplot.}
    \label{fig:s4:ri}
\end{figure}

Figure~\ref{fig:s4:age} presents the fixed effect estimate associated with the age covariate, and Figure~\ref{fig:s4:ri}, the estimated random intercepts for all models.

\paragraph{Missing random slope will mostly affect the corresponding time fixed effect.}

When a non-zero random slope is neglected, the estimates of the age effect (corresponding to the fixed slope or population effect) are centered around the true value, but the coverage is much lower than the nominal level. It is worth mentioning that missing the random slope in the scale model does not influence the fixed effect in the location model, but in the opposite case, the age effect in the scale model shows a loss of coverage. When random intercepts and slopes are independent, neglecting the random slopes induces no changes in the standard deviations of the random intercepts.

\newpage
\subsection{Practice 5: Misspecifying random effect distributions}
Another common practice is to only consider Gaussian distributions for the random intercepts. In this scenario, we explore the misspecification of the random intercept distribution by sampling the random intercept from Student distributions with 3 degrees of freedom instead of a normal distribution, and compare the two following models:
\begin{itemize}
    \item \textbf{Student} correctly specifies the random intercept distributions.
    \begin{verbatim}
          y ~  age + albumin + (1|gr(id, dist=`student'))
          log(omega) ~ age + trig + (1|gr(id, dist=`student'))
    \end{verbatim}
    \item \textbf{Gaussian} assumes Gaussian distributions (the default option).
    \begin{verbatim}
          y ~ age + albumin + (1|id),
          log(omega) ~ age + trig + (1|id)
    \end{verbatim}
\end{itemize}

\begin{figure}[hbtp]
    \centering
    \includegraphics[width=0.49\textwidth]{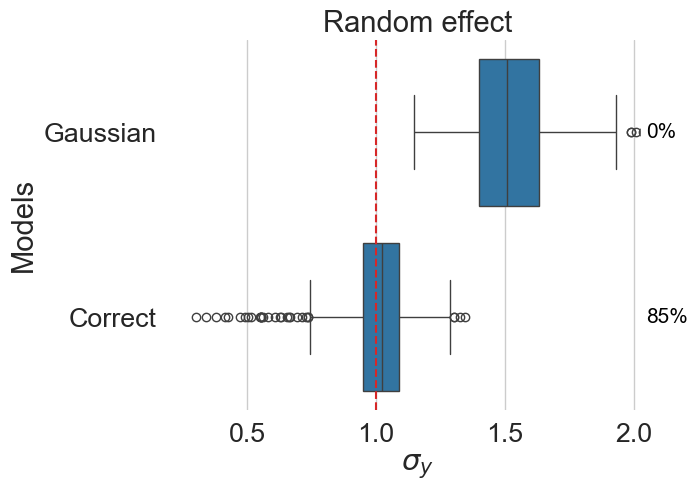}
    \includegraphics[width=0.49\textwidth]{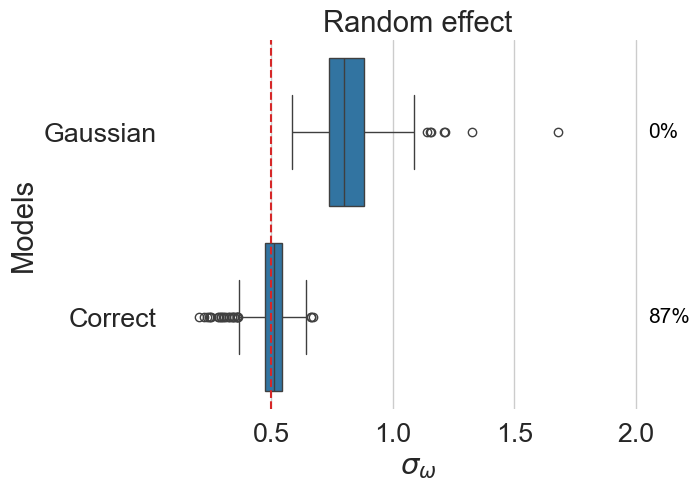}    
    \caption{Estimated standard deviation of random intercepts for location (left panel) and scale (right panel). For each model, coverage is reported as a percentage next to the corresponding boxplot.}
    \label{fig:s5:ri}
\end{figure}

This setting resulted in no difference in the fixed effect estimates; our analysis, therefore, focuses on the random intercepts in Figure~\ref{fig:s5:ri}. 

\begin{figure}[hbtp]
    \centering
    \includegraphics[width=0.4\linewidth]{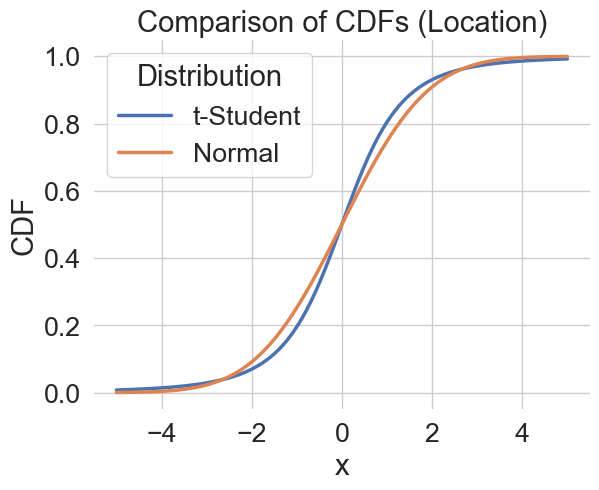}
    \includegraphics[width=0.4\linewidth]{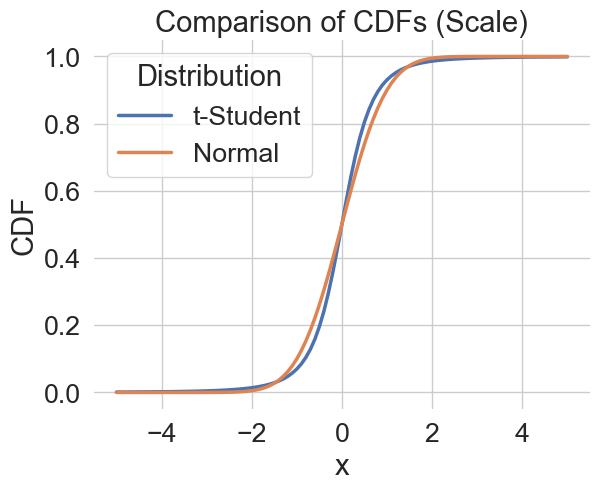}
    \caption{Cumulative distribution function of random effects for the location (left panel) and scale (right panel). The standard deviations of the distributions correspond to the medians in the previous figure.}
    \label{fig:REdistrib}
\end{figure}

\paragraph{Misspecified distributions do not impact on random effect estimation.} 

When the true random effect distribution is t-Student, the standard deviation of the Gaussian distribution will be estimated so that the Gaussian distribution matches the original distribution, as illustrated in Figure~\ref{fig:REdistrib}. This results in similar estimated distributions despite misestimated standard deviations. This result is consistent with the LMM literature in which misspecification of random effects distribution has a limited impact on model estimates~\citep {schielzeth2020robustness}.

\newpage
\section{Case Study: Modelling heteroscedasticity in the PBC dataset}

To further illustrate the impact of these different practices on a study's conclusions, we propose an analysis of the PBC dataset from the R package \textit{JMBayes2}. This example is intended as a didactic demonstration rather than a substantive epidemiological study.

\paragraph{Dataset.} The dataset results from a clinical trial at Mayo Clinic studying the impact of D-penicillamine on primary biliary cirrhosis (PBC), an autoimmune condition affecting the liver. It comprises 1,945 longitudinal observations from 312 patients diagnosed with PBC who received care at the Mayo Clinic. 
These data have been foundational in understanding the progression of this chronic autoimmune liver disease and associated short-term survival~\citep{murtaugh1994primary}. In our analysis, we model log-transformed serum bilirubin levels, a marker of condition severity and progression used for diagnosis and prognosis, as a function of the normalized age at baseline, sex, and time (in years since baseline). 

\paragraph{Empirical setting.} Our goal is to assess how failing to account for heteroscedasticity can affect model estimates. To this end, we compare a traditional LMM and MELSM, which only differ in the way they model the scale. Note that for flexible modelling of the time effect, we use splines (with default parameters as per the package \textit{brms}), denoted by $s(\cdot)$. For fitting these models, we use 4 chains with a burn-in of 1000 iterations for a total of 2000 iterations.
\begin{itemize}

    \item \textbf{LMM} assumes homoscedasticity.
    \begin{verbatim}
        log(bilirubin) ~  age + sex + plat + alb + s(year) + (1|id)
        log(omega) ~ 1
    \end{verbatim}

    \item \textbf{MELSM no covariates} models heteroscedasticity through random effects alone.
    \begin{verbatim}
        log(bilirubin) ~ age + sex + plat + alb + s(year) + (1|id)
        log(omega) ~ (1|id)
    \end{verbatim}

    \item \textbf{MELSM no slope} ignore both random slopes.
    \begin{verbatim}
        log(bilirubin) ~ age + sex + plat + alb + s(year) + (1|id)
        log(omega) ~ age + sex + plat + alb + s(year) + (1|id)
    \end{verbatim}

    \item \textbf{MELSM no slope scale} ignores the scale random slope.
    \begin{verbatim}
        log(bilirubin) ~ age + sex + plat + alb + s(year) + (1 + age|id)
        log(omega) ~ age + sex + plat + alb + s(year) + (1|id)
    \end{verbatim}

    \item \textbf{MELSM slopes} models the location and scale as follows.
    \begin{verbatim}
        log(bilirubin) ~ age + sex + plat + alb + s(year) + (1 + age|id)
        log(omega) ~ age + sex + plat + alb + s(year) + (1 + age|id)
    \end{verbatim}
\end{itemize}

\begin{figure}[ht]
    \centering
    \includegraphics[height=100px]{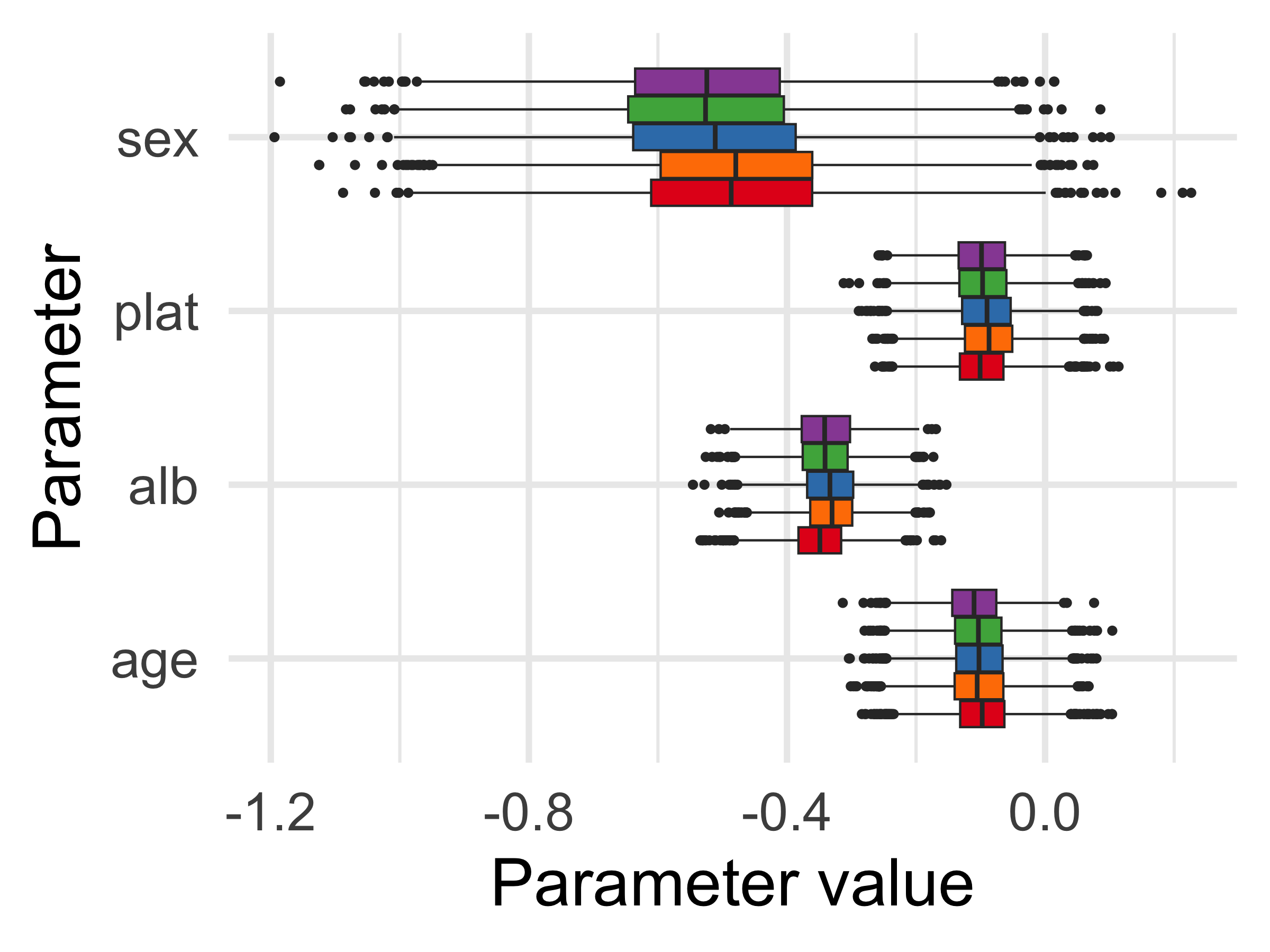}
    \includegraphics[height=100px]{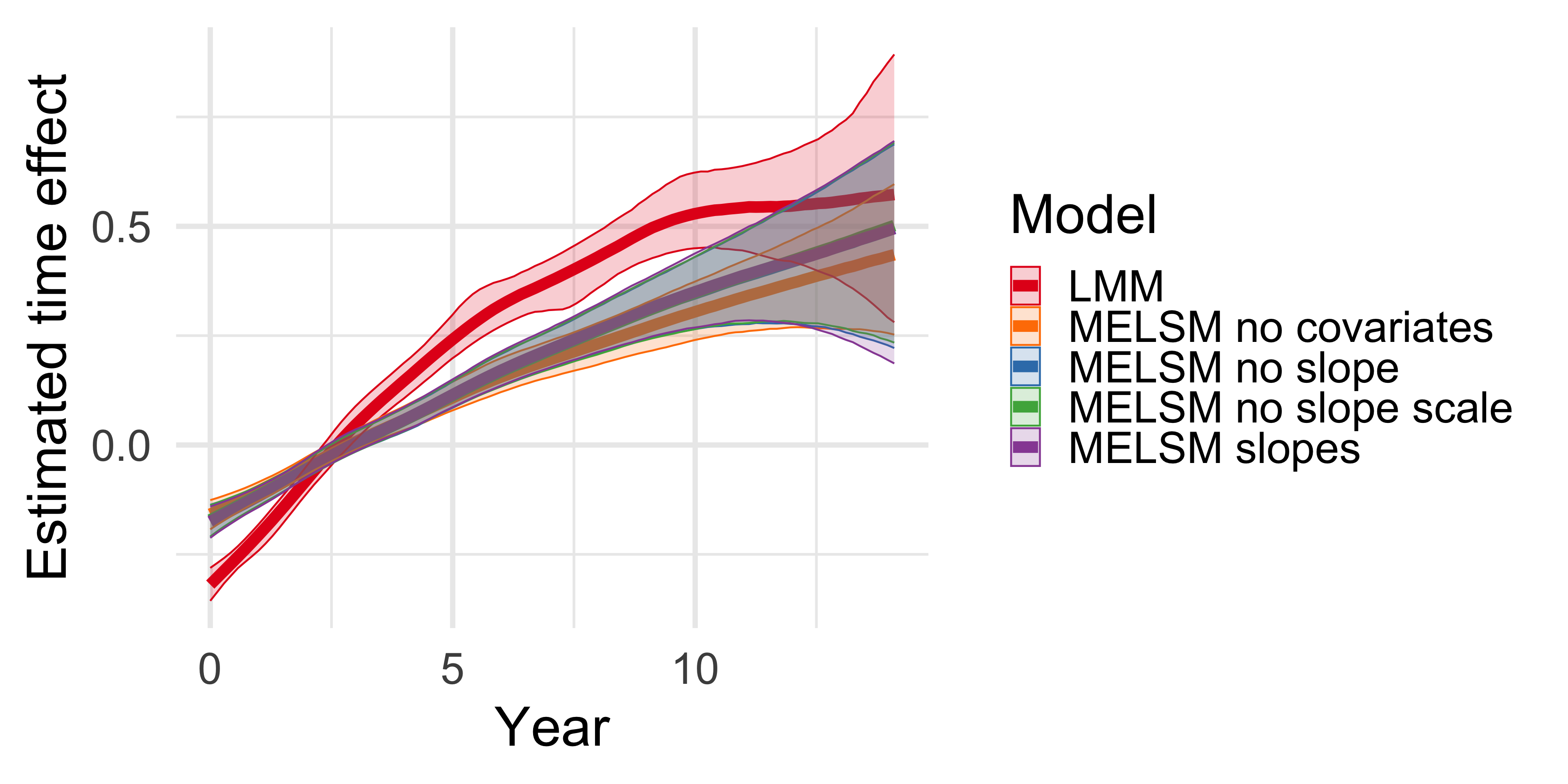}
    \caption{Estimated location fixed effects in the PBC dataset.}
    \label{fig:pbc}
\end{figure}

\paragraph{Results.} Figure~\ref{fig:pbc} displays the fixed effects estimated by all the previous approaches. Critically, there are limited differences in the fixed effects estimated by the different alternatives of MELSM, echoing the finding that misspecification of the scale does not bias fixed effects as long as the heteroscedasticity is captured. While LMM presents similar estimates for the majority of the covariates, the estimated spline for years past since diagnosis differs, presenting a steeper impact. This example illustrates that ignoring error variance can impact inference and underscores the practical importance of explicitly modelling heteroscedasticity when it is present.

\begin{figure}[ht]
    \centering
    \includegraphics[height=100px]{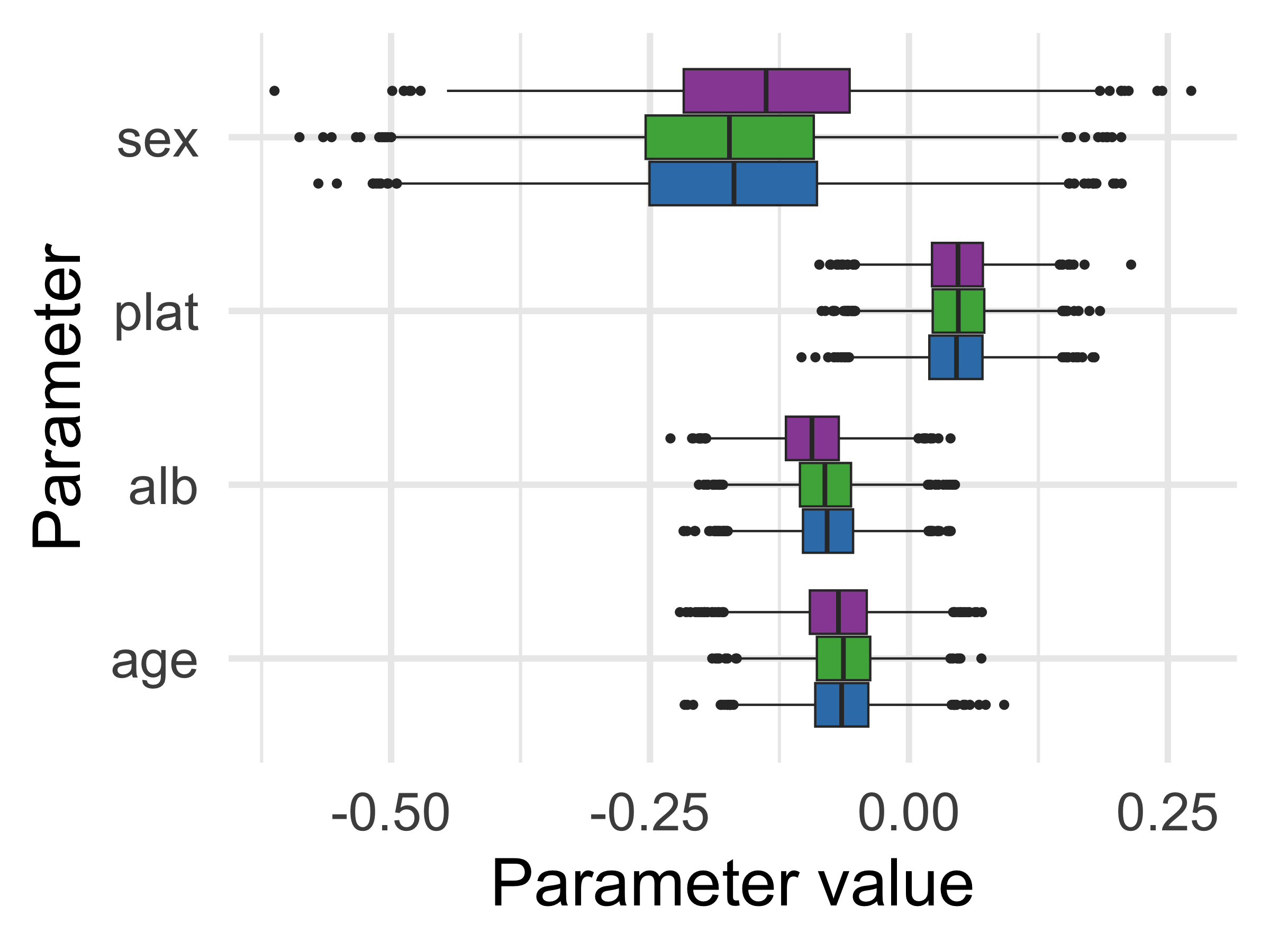}
    \includegraphics[height=100px]{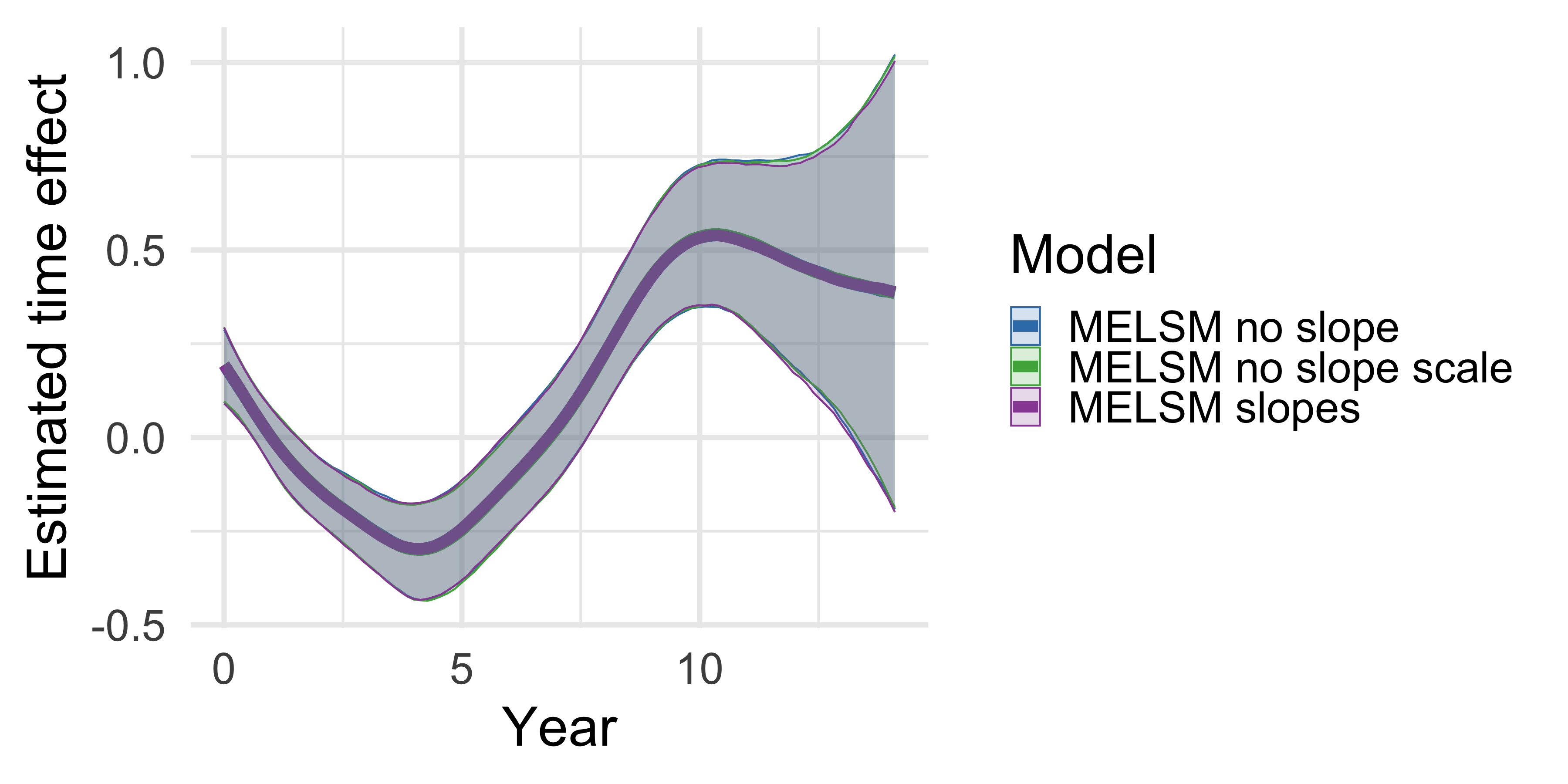}
    \caption{Estimated scale fixed effects in the PBC dataset.}
    \label{fig:scale}
\end{figure}

Figure~\ref{fig:scale} presents the scale fixed effects for the three models considering the different covariates in the scale model. The estimates are similar across models, indicating a limited impact of introducing random slopes in the specification of the scale submodel.

\begin{figure}[ht]
    \centering
    \includegraphics[height=100px]{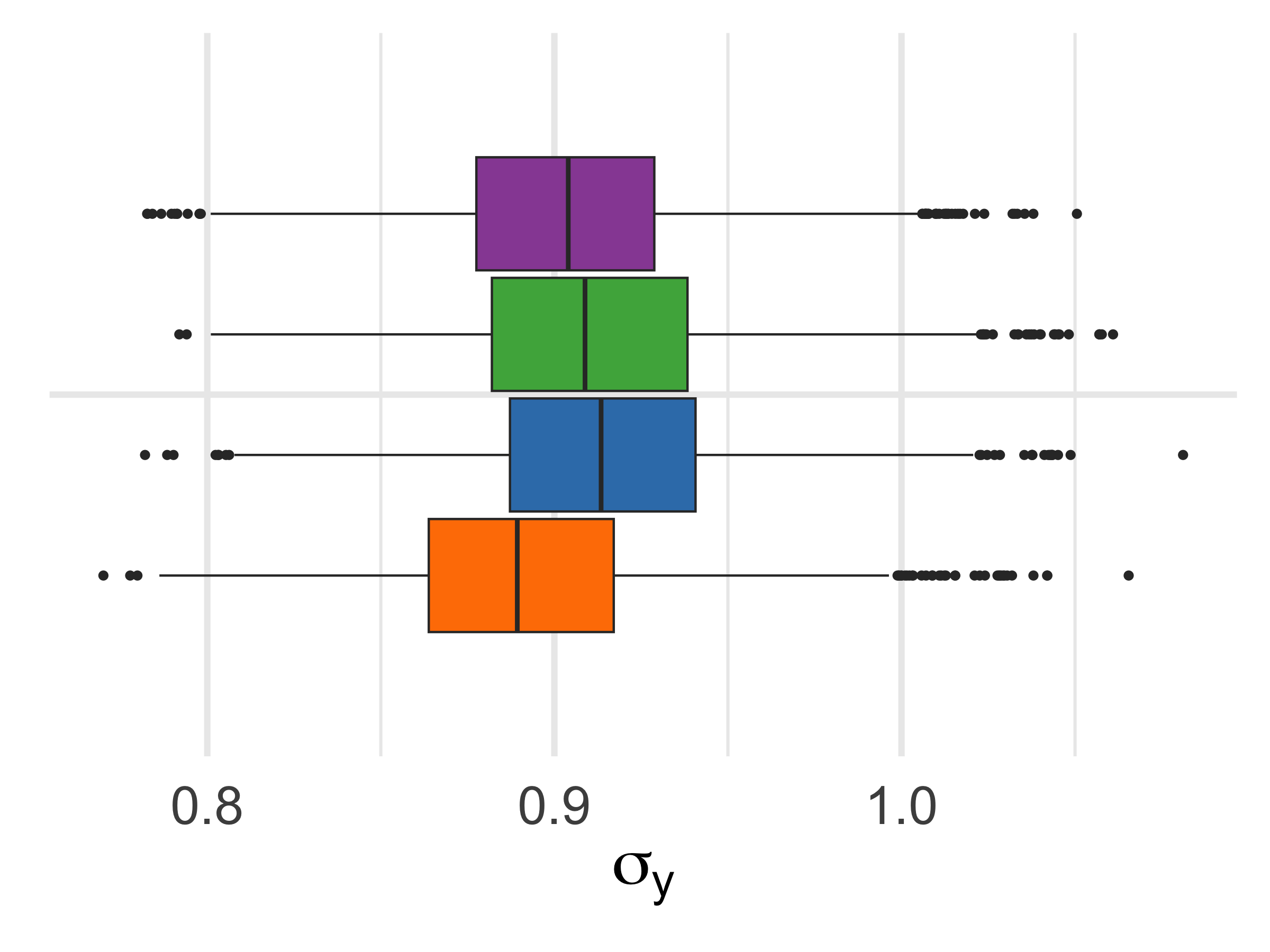}
    \includegraphics[height=100px]{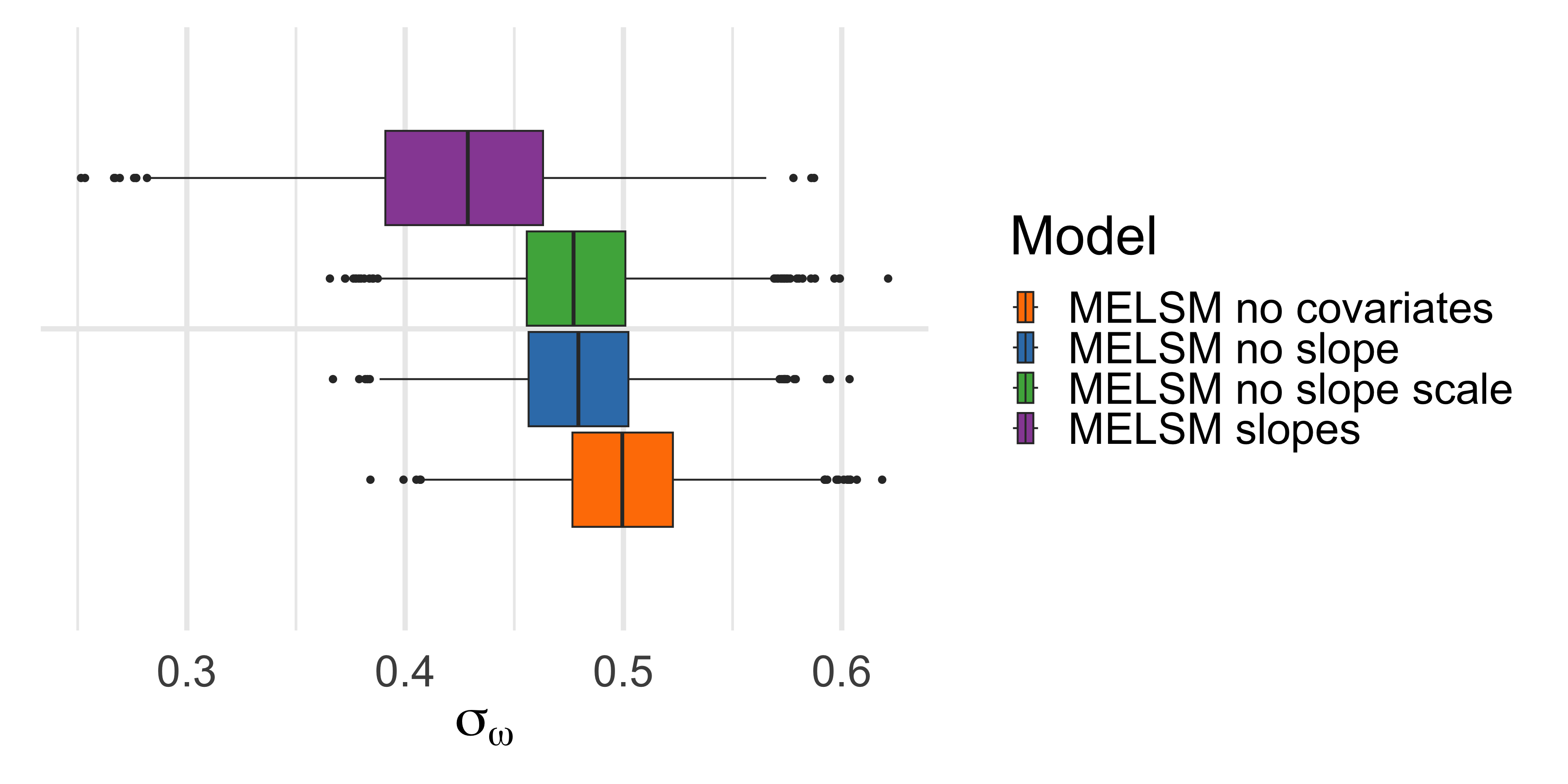}
    \caption{Estimated standard deviation of random intercepts for location (left panel) and scale (right panel) in the PBC dataset.}
    \label{fig:deviation}
\end{figure}

Figure~\ref{fig:deviation} presents the estimated standard deviation of the random effects for all MELSM models considering heteroscedasticity. The standard deviation of the scale random effects presents larger values for MELSM with no covariates, as this model does not control for any other covariates. Once covariates or random slopes are accounted for in the scale submodel, the standard deviation of the estimated random intercepts decreases. Importantly, note that the posterior distribution for the scale random intercept lies above 0, providing therefore evidence of heteroscedasticity.

\section{Discussion}

In this work, we present a set of simulations to study the impact of misspecification on mixed-effect location-scale models, with a focus on modelling heteroscedasticity as opposed to the assumption of constant residual variance made by standard linear mixed models.

Our simulations offer valuable insights into MELSM. First, scale misspecification increases the variance of the location estimates but does not introduce bias. In contrast, location misspecification biases scale estimates. Therefore, all conclusions about covariate effects in the scale model are to be referred to the specific set of covariates and functional forms used in the location. 
Further, there is an interplay between the random effects in the location model and the residual variance. Different models for residual variance influence the standard deviations of the location random effects. 
Finally, misspecification of the model structure, such as neglecting a relevant covariate, leads to larger bias than misspecification of the functional form, as including relevant covariates helps account for part of their effects, even if the functional form is not correct.
These insights highlight the critical importance of modelling residual variance, inviting statisticians to model this quantity even under imperfect knowledge of its structure.

As a simulation, our findings have limitations. While we have used a general set of parameters to improve the generalisability of our conclusions, more extreme parameter settings could depart from the results shown here. 
Similarly, our simulations rely on two random effects and do not account for random effects in the B-spline coefficients.
Our conclusion regarding the misspecification of the random effect distribution considers a symmetric alternative to the normal distribution, rather than more skewed distributions, which could impact this finding. 
Further, the proposed simulations assume an uninformative visit process, which in many real cases (e.g.\ electronic health records) might be unrealistic. Future developments in this area are needed to build joint models for the within-individual variability and the informative observation process, extending, for example, the methodologies illustrated in \citet{pullenayegum2016longitudinal} and \citet{garrett2024biasmixedmodelsanalysing}.
Finally, our work focuses on the distribution shape, rather than the prediction of individual random effects, which the linear mixed models literature suggests is more affected by misspecification (see \citealp{lebeau2018model, hui2021random, vu2025random} for reviews on the topic).

Our findings may have implications for other longitudinal frameworks dependent on MELSM. 
Beyond the presented setting, MELSMs are often used as part of more complex models. Particularly, the study of joint models with within-individual variability~\citep{barrett2019estimating, palma2025bayesian} would provide insights into the relationship between the misspecification of the longitudinal model and the association with the hazard in time-to-event data analysis.

\newpage
\backmatter
\section*{Declarations}
For the purpose of open access, the author has applied a Creative Commons Attribution (CC BY) licence to any Author Accepted Manuscript version arising from this submission.\\
This version of the article has been accepted for publication, after peer review, but is not the Version of Record and does not reflect post-acceptance improvements or any corrections. The Version of Record is available online at:  http://dx.doi.org/10.1186/s12874-025-02755-3.\\
\textbf{Ethics approval and consent to participate.} Not applicable.\\
\textbf{Consent for publication.} Not applicable.\\
\textbf{Availability of data and materials.} All data generated and code to reproduce the current study are available at~\url{https://github.com/Jeanselme/IndividualVariability}.\\
\textbf{Competing interests.} JKB has received research funding for unrelated work from F. Hoffmann-La Roche Ltd.\\
\textbf{Funding.} This work was supported by the UK Medical Research Council grant “Looking beyond the mean: what within-person variability can tell us about dementia, cardiovascular disease and cystic fibrosis” (MR/V020595/1). JKB was supported through the United Kingdom Medical Research Council programme (grant MC\_UU\_00002/5) and theme (grant MC\_UU\_00040/02 – Precision Medicine) funding. MP's research is currently supported by the Ulverscroft Vision Research Group (UCL).\\
\textbf{Authors' contributions.} VJ: Formal analysis, software, writing—original draft. JKB: Funding acquisition. MP, JKB: Supervision. VJ, MP, JKB: Conceptualization, writing—reviewing and editing.\\
\textbf{Acknowledgements.} The authors would like to thank the reviewers for their valuable feedback throughout the review process.

\bibliography{sn-bibliography}

\begin{appendices}

\section{Why do we expect a biased location model to have a larger scale error?}
\label{app:location-scale}
Let us simplify the problem by considering the following generative process: $$Y = \beta X + \epsilon$$ with $\epsilon \sim \mathcal{N}(0, \omega)$.

A biased scale model can be expressed as: $$Y = (\beta + \delta) X + \epsilon'$$ with $\epsilon' \sim \mathcal{N}(0, \omega')$ and $\delta$ the bias from the true coefficient.

Computing the difference between the generative process and the biased model results in:
$$\epsilon' = \epsilon - \delta X$$ with $X \sim MVN(0, \Sigma)$. And therefore, one can express the variance relation as:
$$\omega' = \omega + \delta^T \Sigma \delta$$
Note that the exact quantification of these quantities is dependent upon the correlation structure between covariates. However, as $\Sigma$ is positive semi-definite: $$\omega' \geq \omega$$
This means that the estimated variance of a biased model is larger than the variance of the data.

\section{Additional metrics}
\label{app:metrics}
In addition to the coverage and estimates presented in the main text, we present the additional metrics of the average credible interval width ($\bar{CI}(\theta)$), the standard deviation of the estimates $SD(\theta)$, and average posterior standard deviations $\bar{\epsilon}(\theta)$ for the same parameters of interest across the different replications.

\subsection{Practice 1: Ignoring heteroscedasticity}
These tables echo the same conclusions as the main text, with the correctly specified model presenting smaller credible interval widths, deviations of the estimate, and average standard errors by an order of magnitude. Increasing numbers of individuals or points also result in improvements across all metrics. 
\begin{table}[!h]
    \centering
    \begin{tabular}{ccccc}
\toprule
 &  & $SD$($\beta^y[Age]$) & $\bar{CI}$($\beta^y[Age]$) & $\bar{\epsilon}$($\beta^y[Age]$) \\
 & Models &  &  &  \\
\midrule
100 & Correct & 0.017 & 0.063 & 0.016 \\
\midrule
300 & Correct & 0.009 & 0.035 & 0.009 \\
\midrule
500 & Correct & 0.007 & 0.026 & 0.007 \\
\midrule
1000 & Correct & 0.005 & 0.018 & 0.005 \\
\bottomrule
\end{tabular}

    \caption{Measures of errors of age effect on location for increasing number of individuals.}
\end{table}

\begin{table}[!h]
    \centering
    \begin{tabular}{ccccc}
\toprule
 &  & $SD$($\beta^{\omega}[Age]$) & $\bar{CI}$($\beta^{\omega}[Age]$) & $\bar{\epsilon}$($\beta^{\omega}[Age]$) \\
 & Models &  &  &  \\
\midrule
100 & Correct & 0.024 & 0.092 & 0.023 \\
\midrule
300 & Correct & 0.013 & 0.053 & 0.013 \\
\midrule
500 & Correct & 0.010 & 0.041 & 0.010 \\
\midrule
1000 & Correct & 0.008 & 0.029 & 0.007 \\
\bottomrule
\end{tabular}

    \caption{Measures of errors of age effect on scale for increasing number of individuals.}
\end{table}

\begin{table}[!h]
    \centering
    \begin{tabular}{ccccc}
\toprule
 &  & $SD$($\beta^y[Age]$) & $\bar{CI}$($\beta^y[Age]$) & $\bar{\epsilon}$($\beta^y[Age]$) \\
 & Models &  &  &  \\
\midrule
5 & Correct & 0.021 & 0.082 & 0.021 \\
\midrule
10 & Correct & 0.013 & 0.055 & 0.014 \\
\midrule
20 & Correct & 0.009 & 0.037 & 0.009 \\
\bottomrule
\end{tabular}
  
    \caption{Measures of errors of age effect on location for increasing number of points.}
\end{table}

\begin{table}[!h]
    \centering
    \begin{tabular}{ccccc}
\toprule
 &  & $SD$($\beta^{\omega}[Age]$) & $\bar{CI}$($\beta^{\omega}[Age]$) & $\bar{\epsilon}$($\beta^{\omega}[Age]$) \\
 & Models &  &  &  \\
\midrule
5 & Correct & 0.039 & 0.140 & 0.036 \\
\midrule
10 & Correct & 0.022 & 0.084 & 0.021 \\
\midrule
20 & Correct & 0.014 & 0.055 & 0.014 \\
\bottomrule
\end{tabular}
    
    \caption{Measures of errors of age effect on scale for increasing number of points.}
\end{table}

\clearpage
\subsection{Practice 2: Misspecifying location and scale in MELSM}
Similarly, this section illustrates how different misspecifications affect the different fixed effects with regard to these metrics. Interestingly, the model that considers all covariates and the correctly specified one are indistinguishable from these metrics. 
\begin{table}[!h]
    \centering
    \begin{tabular}{cccc}
\toprule
 & $SD$($\beta^y[Age]$) & $\bar{CI}$($\beta^y[Age]$) & $\bar{\epsilon}$($\beta^y[Age]$) \\
Models &  &  &  \\
\midrule
Correct & 0.011 & 0.043 & 0.011 \\
All & 0.011 & 0.043 & 0.011 \\
Mis. $\omega$ & 0.018 & 0.056 & 0.014 \\
No $u^\omega$ & 0.017 & 0.065 & 0.017 \\
No $u^\omega$ + Mis. $\omega$ & 0.044 & 0.109 & 0.028 \\
\bottomrule
\end{tabular}
  
    \caption{Measures of errors of age effect on location.}
\end{table}

\begin{table}[!h]
    \centering
    \begin{tabular}{cccc}
\toprule
 & $SD$($\beta^{\omega}[Age]$) & $\bar{CI}$($\beta^{\omega}[Age]$) & $\bar{\epsilon}$($\beta^{\omega}[Age]$) \\
Models &  &  &  \\
\midrule
Correct & 0.016 & 0.065 & 0.017 \\
All & 0.016 & 0.065 & 0.017 \\
Mis. $y$ & 0.017 & 0.068 & 0.017 \\
No $u^\omega$ & 0.032 & 0.055 & 0.014 \\
\bottomrule
\end{tabular}
    
    \caption{Measures of errors of age effect on scale.}
\end{table}

\begin{table}[!h]
    \centering
    \begin{tabular}{cccc}
\toprule
 & $SD$($\sigma_{y}$) & $\bar{CI}$($\sigma_{y}$) & $\bar{\epsilon}$($\sigma_{y}$) \\
Models &  &  &  \\
\midrule
Correct & 0.054 & 0.217 & 0.055 \\
All & 0.054 & 0.218 & 0.056 \\
Mis. $y$ & 0.054 & 0.218 & 0.056 \\
Mis. $\omega$ & 0.058 & 0.234 & 0.060 \\
No $u^\omega$ & 0.055 & 0.219 & 0.056 \\
No $u^\omega$ + Mis. $\omega$ & 0.071 & 0.266 & 0.068 \\
\bottomrule
\end{tabular}
  
    \caption{Measures of errors of standard deviation of random intercepts for location.}
\end{table}

\begin{table}[!h]
    \centering
    \begin{tabular}{cccc}
\toprule
 & $SD$($\sigma_{\omega}$) & $\bar{CI}$($\sigma_{\omega}$) & $\bar{\epsilon}$($\sigma_{\omega}$) \\
Models &  &  &  \\
\midrule
Correct & 0.031 & 0.123 & 0.031 \\
All & 0.031 & 0.124 & 0.032 \\
Mis. $y$ & 0.029 & 0.109 & 0.028 \\
Mis. $\omega$ & 0.041 & 0.185 & 0.047 \\
\bottomrule
\end{tabular}
    
    \caption{Measures of errors of standard deviation of random intercepts for scale.}
\end{table}

\clearpage
\subsection{Practice 3: Ignoring non-linear time trend in location model}
The following tables illustrate these metrics when the correctly specified model depends on the sine of age. Interestingly, these metrics do not capture the bias in the estimated fixed effect, even if the improperly specified model shows narrower parameter estimates. This aligns with the intuition that more complex models may lead to larger standard deviations for the different parameters of interest.
\begin{table}[!h]
    \centering
    \begin{tabular}{cccc}
\toprule
 & $SD$($\beta^y[Age]$) & $\bar{CI}$($\beta^y[Age]$) & $\bar{\epsilon}$($\beta^y[Age]$) \\
Models &  &  &  \\
\midrule
Correct & 0.015 & 0.058 & 0.015 \\
Non sinus & 0.014 & 0.047 & 0.012 \\
\bottomrule
\end{tabular}
  
    \caption{Measures of errors of age effect on location.}
\end{table}

\begin{table}[!h]
    \centering
    \begin{tabular}{cccc}
\toprule
 & $SD$($\beta^{\omega}[Age]$) & $\bar{CI}$($\beta^{\omega}[Age]$) & $\bar{\epsilon}$($\beta^{\omega}[Age]$) \\
Models &  &  &  \\
\midrule
Correct & 0.016 & 0.065 & 0.017 \\
Non sinus & 0.016 & 0.065 & 0.017 \\
\bottomrule
\end{tabular}
    
    \caption{Measures of errors of age effect on scale.}
\end{table}

\clearpage
\subsection{Practice 4: Misspecifying the random effect structure}
In this setting, the metrics again do not capture the estimate bias, as the correctly specified model presents worse performance on the random effect estimates despite presenting less bias.

The tables show the effect of removing the random slope for age in each of the two submodels. We observe that, once we remove the random slope coefficient from the location submodel, the corresponding fixed effect is affected by low coverage and the estimates of the metrics in the tables are different from those in the correct model. A similar behaviour is observed for the random slope in the scale submodel. 

The metrics for the standard deviation of the random intercept in the location and scale submodels remains essentially unaltered after removing the random slopes.  
\begin{table}[!h]
    \centering
    \begin{tabular}{cccc}
\toprule
 & $SD$($\beta^y[Age]$) & $\bar{CI}$($\beta^y[Age]$) & $\bar{\epsilon}$($\beta^y[Age]$) \\
Models &  &  &  \\
\midrule
Correct & 0.031 & 0.120 & 0.031 \\
No $u^\omega_{age}$ & 0.032 & 0.121 & 0.031 \\
No slopes & 0.046 & 0.066 & 0.017 \\
\bottomrule
\end{tabular}
  
    \caption{Measures of errors of age effect on location.}
\end{table}

\begin{table}[!h]
    \centering
    \begin{tabular}{cccc}
\toprule
 & $SD$($\beta^{\omega}[Age]$) & $\bar{CI}$($\beta^{\omega}[Age]$) & $\bar{\epsilon}$($\beta^{\omega}[Age]$) \\
Models &  &  &  \\
\midrule
Correct & 0.029 & 0.114 & 0.029 \\
No $u^\omega_{age}$ & 0.033 & 0.068 & 0.017 \\
No slopes & 0.032 & 0.068 & 0.017 \\
\bottomrule
\end{tabular}
    
    \caption{Measures of errors of age effect on scale.}
\end{table}

\begin{table}[!h]
    \centering
    \begin{tabular}{cccc}
\toprule
 & $SD$($\sigma_{y}$) & $\bar{CI}$($\sigma_{y}$) & $\bar{\epsilon}$($\sigma_{y}$) \\
Models &  &  &  \\
\midrule
Correct & 0.060 & 0.236 & 0.060 \\
No $u^\omega_{age}$ & 0.060 & 0.237 & 0.061 \\
No slopes & 0.053 & 0.223 & 0.057 \\
\bottomrule
\end{tabular}
  
    \caption{Measures of errors of standard deviation of random intercepts for location.}
\end{table}

\begin{table}[!h]
    \centering
    \begin{tabular}{cccc}
\toprule
 & $SD$($\sigma_{\omega}$) & $\bar{CI}$($\sigma_{\omega}$) & $\bar{\epsilon}$($\sigma_{\omega}$) \\
Models &  &  &  \\
\midrule
Correct & 0.034 & 0.132 & 0.034 \\
No $u^\omega_{age}$ & 0.034 & 0.129 & 0.033 \\
No slopes & 0.032 & 0.125 & 0.032 \\
\bottomrule
\end{tabular}
    
    \caption{Measures of errors of standard deviation of random intercepts for scale.}
\end{table}

\clearpage
\subsection{Practice 5: Misspecifying random effect distributions}
In this final experiment, the proposed metrics reveal limited differences between the two distribution specifications of the random effects (except for $SD(\sigma_y)$, which is dependent on the different distribution parameterisation). This is consistent with the findings on the limited impact of distribution misspecification.
\begin{table}[!h]
    \centering
    \begin{tabular}{cccc}
\toprule
 & $SD$($\sigma_{y}$) & $\bar{CI}$($\sigma_{y}$) & $\bar{\epsilon}$($\sigma_{y}$) \\
Models &  &  &  \\
\midrule
Correct & 0.147 & 0.345 & 0.093 \\
Gaussian & 0.190 & 0.329 & 0.086 \\
\bottomrule
\end{tabular}
  
    \caption{Measures of errors of standard deviation of random intercepts for location.}
\end{table}

\begin{table}[!h]
    \centering
    \begin{tabular}{cccc}
\toprule
 & $SD$($\sigma_{\omega}$) & $\bar{CI}$($\sigma_{\omega}$) & $\bar{\epsilon}$($\sigma_{\omega}$) \\
Models &  &  &  \\
\midrule
Correct & 0.068 & 0.190 & 0.051 \\
Gaussian & 0.142 & 0.182 & 0.047 \\
\bottomrule
\end{tabular}
    
    \caption{Measures of errors of standard deviation of random intercepts for scale.}
\end{table}

\end{appendices}



\end{document}